\documentclass[12pt, a4paper]{article}
\usepackage{amsfonts,amssymb,amsmath,amsthm,graphicx}

\newcommand{\BB}{\vert_{\partial M}}
\newcommand{\tr}{{\rm tr}\,}
\newcommand{\Dir}{\widehat D}
\newcommand{\iM}{\int_M d^nx\sqrt{g}}
\newcommand{\idM}{\int_{\partial M} d^{n-1}x\sqrt{h}}

\begin{document}

\title{One loop boundary effects:
techniques and applications. \thanks{Lectures given at International
V.A.Fock School for Advances of Physics IFSAP-2005, St.Petersburg,
Russia, November 21-27, 2005.} }

\author{Valery N.Marachevsky \thanks{email: maraval@mail.ru} \\
{\it V. A. Fock Institute of Physics, St. Petersburg
University,}\\
{\it 198504 St. Petersburg, Russia} }

%\author{<author2>}{
%  address={<common address for author2 and author3>}
%}

%\author{<author3>}{
%  address={<common address for author2 and author3>}
%  ,altaddress={<author1 address>} % additional visiting address
%}

\maketitle

\begin{abstract}
A pedagogical introduction to the heat kernel technique, zeta
function and Casimir effect is presented. Several applications are
considered. First we derive the high temperature asymptotics of the
free energy for boson fields in terms of the heat kernel expansion
and zeta function. Another application is chiral anomaly for local
(MIT bag) boundary conditions. Then we rederive the Casimir energies
for perfectly conducting rectangular cavities using a new technique.
The new results for the attractive Casimir force acting on each of
the two perfectly conducting plates inside an infinite perfectly
conducting waveguide of the same cross section as the plates are
presented at zero and finite temperatures.
\end{abstract}

\maketitle

%%%%%%%%%%%%%%%%%%%%%%%%%%%%%%%%%%%%%%%%%%%%
%% MAINMATTER
%%%%%%%%%%%%%%%%%%%%%%%%%%%%%%%%%%%%%%%%%%%%
\section{Introduction}

The main problem of the quantum field theory with boundaries is its
renormalization and physical meaning of the results obtained.
Divergences that appear in quantum field theory make the problems on
manifolds with boundaries more complicated than in infinite space.

In the presence of boundaries or singularities the heat kernel
technique is an effective tool for the analysis of the one loop
effects (see reviews \cite{Vassilevich:2003xt}, \cite{Santangelo2}).
Different applications of the heat kernel expansion exist. The heat
kernel technique seems to be the easiest way for the calculation of
quantum anomalies, calculation of effective actions based on
finite-mode regularization and analysis of divergences in quantum
field theory.

Chiral anomaly, which was discovered more than 35 years ago
\cite{abj}, still plays an important role in physics. On smooth
manifolds without boundaries many successful approaches to the
anomalies exist \cite{Jackiw},\cite{Bertlmann}, \cite{Fujikawa1}.
The heat kernel approach to the anomalies is essentially equivalent
to the Fujikawa approach \cite{Fujikawa:1979ay} and to the
calculations based on the finite-mode regularization
\cite{Andrianov:1983fg}, but it can be more easily extended to
complicated geometries. The local chiral anomaly in the case of
non-trivial boundary conditions (MIT bag boundary conditions) has
been calculated only recently \cite{lastpaper}.

Casimir effect \cite{Casimir} is a macroscopic quantum effect.
Briefly speaking, if we impose classical boundary conditions on a
quantum field on some boundary surface  than we get the Casimir
effect. There are several different physical situations that should
be distinguished in the Casimir effect.

Suppose there are two spatially separated dielectrics, then in a
dilute limit ($\epsilon \to 1$) the Casimir energy of this system
is equal to the energy of pairwise interactions between dipoles of
these two dielectrics via a Casimir-Polder retarded potential
\cite{Polder}. For a general case of separated dielectrics the
Casimir energy can be calculated as in \cite{Lifshitz} or
\cite{Milton} (for a recent discussion of these issues see
\cite{Brevik} and a review \cite{Milton2}, new possible
experiments in \cite{Giampiero} ).

A different situation takes place when there is a dilute dielectric
ball or any other simply connected dielectric under study (see a
review \cite{Nesterenko} for a discussion of related subjects and
methods used). As it was pointed out in \cite{Marachevsky} and then
discussed in detail in \cite{Marachevsky2}, microscopic interatomic
distances should be taken into account to calculate the Casimir
energy  of a dilute dielectric ball. The average interatomic
distance $\lambda$ serves as an effective physical cut off for
simply connected dielectrics.

The limit of a perfect conductivity ($\epsilon \to +\infty$) is
opposite to a dilute case. This is the strong coupling limit of
the theory. Any results obtained in this limit are nonperturbative
ones.

The Casimir energy for a perfectly conducting rectangular cavity
was first calculated in \cite{Lukosz} using exponential
regularization. Later it was derived by some other methods (see
references and numerical analysis in \cite{Maclay}, also a review
\cite{Bordag1}). In the present paper we derive the Casimir energy
for rectangular cavities at zero temperature by a new method
described in Sec.$3$. By use of this method we could rewrite the
Casimir energy for rectangular cavities in the form that makes
transparent its geometric interpretation. Also this method yields
new exact results for the Casimir force acting on two or more
perfectly conducting plates of an arbitrary cross section inside
an infinite perfectly conducting waveguide of the same cross
section.

The paper is organized as follows. In  Sec.$2$ we give an
introduction to the formalism of the heat kernel and heat kernel
expansion. Also we introduce a $\zeta$-function \cite{Odintsov}
and calculate the one loop effective action in terms of $\zeta$-
function. Then we consider two examples. First we derive the high
temperature expansion of the free energy for boson fields
\cite{Dowker} in terms of the heat kernel expansion and
$\zeta$-function. Then we derive a chiral anomaly in four
dimensions for an euclidean version of the MIT bag boundary
conditions \cite{bag}. Sec.$3$ is devoted to the Casimir effect
for perfectly conducting cavities. In Sec.$3.1$ we introduce a
regularization and a convenient new method of calculations using
an example of two perfectly conducting parallel plates, then apply
it to more complicated rectangular geometries. We present the
Casimir energy of the cavity in the form (\ref{box}). Then we
discuss an argument principle and $\zeta$-functional
regularization for the cavity. In Sec.$3.6$ we rewrite the Casimir
energy of the cavity in terms of geometric optics (\ref{main}). In
Sec.$3.7$ we describe a possible experiment  and derive the
formula for the attractive force acting on each of the two
parallel plates inside an infinite rectangular waveguide with the
same cross section. Also we present a generalization of this
result for the case of the two parallel perfectly conducting
plates of an arbitrary cross section inside an infinite perfectly
conducting waveguide with the same cross section at zero and
finite temperatures.

\section{Spectral techniques}
\subsection{Heat kernel}

Consider a second order elliptic partial differential operator $L$
of Laplace type on an n-dimensional Riemannian manifold. Any
operator of this type can be expanded locally as
\begin{equation}
  L = - (g^{\mu\nu} \partial_\mu \partial_\nu + a^\sigma \partial_\sigma + b ),
  \label{B4}
\end{equation}
where $a$ and $b$ are some matrix valued functions and $g^{\mu\nu}$
is the inverse metric tensor on the manifold. For a flat space
$g^{\mu\nu}=\delta^{\mu\nu}$.

The heat kernel can be defined as follows:
\begin{equation}
K(t; x; y; L) = \langle x | \exp (-tL) | y \rangle = \sum_{\lambda}
\phi_{\lambda}^{\dagger} (x) \phi_{\lambda} (y) \exp (-t\lambda)  ,
\end{equation}
where $\phi_{\lambda}$ is an eigenfunction of the operator $L$ with
the eigenvalue $\lambda$.

It satisfies the heat equation
\begin{equation}
(\partial_t + L_x) K (t; x ; y ; L) = 0
\end{equation}
with an initial condition
\begin{equation}
K(0; x; y; L) = \delta (x, y) .
\end{equation}

If we consider the fields in a finite volume then it is necessary to
specify boundary conditions. Different choices are possible. In
section $3.1$ we will consider the case of periodic boundary
conditions on imaginary time coordinate, which are specific for
boson fields. In section $3.2$ we will study bag boundary conditions
imposed on fermion fields. If the normal to the boundary component
of the fermion current $\psi^\dag \gamma_n \psi$ vanishes at the
boundary, one can impose bag boundary conditions, a particular case
of mixed boundary conditions. We assume given two complementary
projectors $\Pi_\pm$, $\Pi_-+\Pi_+=I$ acting on a multi component
field (the eigenfunction of the operator $L$) at each point of the
boundary and define mixed boundary conditions by the relations
\begin{equation}
\Pi_-\psi \BB =0\,,\quad \left( \nabla_n + S\right) \Pi_+ \psi \BB
=0 \,, \label{mixedbc}
\end{equation}
where $S$ is a matrix valued function on the boundary. In other
words, the components $\Pi_-\psi$ satisfy Dirichlet boundary
conditions, and $\Pi_+\psi$ satisfy Robin (modified Neumann) ones.

It is convenient to define
\begin{equation}
\chi =\Pi_+ - \Pi_- \,.\label{defchi}
\end{equation}

Let $\{ e_j \}$, $j=1,\dots,n$ be a local orthonormal frame for the
tangent space to the manifold and let on the boundary $e_n$ be an
inward pointing normal vector.

The extrinsic curvature is defined by the equation
\begin{equation}
L_{ab}=\Gamma_{ab}^n \,,\label{Lab}
\end{equation}
where $\Gamma$ is the Christoffel symbol. For example, on the unit
sphere $S^{n-1}$ which bounds the unit ball in $R^n$ the extrinsic
curvature is $L_{ab}=\delta_{ab}$.

Curved space offers no complications in our approach compared to the
flat case. Let $R_{\mu\nu\rho\sigma}$ be the Riemann tensor, and let
$R_{\mu\nu}={R^\sigma}_{\mu\nu\sigma}$ be the Ricci tensor. With our
sign convention the scalar curvature $R=R_\mu^\mu$ is $+2$ on the
unit sphere $S^2$. In flat space the Riemann and Ricci tensors are
equal to zero.

One can always introduce a connection $\omega_\mu$ and another
matrix valued function $E $ so that $L$ takes the form:
\begin{equation}
  L= - (g^{\mu\nu}\nabla_\mu \nabla_\nu + E)  \label{B5}
\end{equation}
Here $\nabla_\mu $ is a sum of covariant Riemannian derivative with
respect to metric $g_{\mu\nu}$ and connection $\omega_\mu$. One can,
of course, express $E$ and $\omega$ in terms of $a^\mu$, $b$ and
$g_{\mu\nu}$:

\begin{eqnarray}
&&\omega_\mu = \frac{1}{2} g_{\mu\nu} (a^\nu +
  g^{\rho\sigma}\Gamma_{\rho\sigma}^\nu) , \label{B6}
\\
&& E = b - g^{\mu\nu}(\partial_\nu \omega_\mu + \omega_\mu\omega_\nu
  -\omega_\rho\Gamma_{\mu\nu}^\rho )    \label{B7}
\end{eqnarray}

For the future use we introduce also the field strength for
$\omega$:
\begin{equation}
 \Omega_{\mu\nu}=\partial_\mu\omega_\nu -
\partial_\nu\omega_\mu + [\omega_\mu, \omega_\nu] \,.\label{Omega}
\end{equation}

The connection $\omega_\mu$ will be used to construct covariant
derivatives.  The subscript $;\mu\dots \nu\sigma$ will be used to
denote repeated covariant derivatives with the connection $\omega$
and the Christoffel connection on $M$. The subscript $:a\dots b c$
will denote repeated covariant derivatives containing $\omega$ and
the Christoffel connection on the boundary. Difference between these
two covariant derivatives is measured by the extrinsic curvature
(\ref{Lab}). For example, $E_{;ab}=E_{:ab}-L_{ab}E_{;n}$.

Let us define an integrated heat kernel for a hermitian operator $L$
by the equation:
\begin{equation}
K(Q,L,t):={\rm Tr} \left( Q \exp (-tL) \right) = \iM \tr \left( Q(x)
K(t;x;x;L) \right) \,,\label{defhk}
\end{equation}
where $Q(x)$ is an hermitian matrix valued function, ${\rm tr}$ here
is over matrix indices. For the boundary conditions we consider in
this paper there exists an asymptotic expansion \cite{Gilkey:1994}
as $t\to 0$:
\begin{equation}
K(Q,L,t) \simeq \sum_{k=0}^\infty a_k (Q,L) t^{(k-n)/2} \,.
\label{hkexp}
\end{equation}

According to the general theory \cite{Gilkey:1994} the coefficients
$a_k(Q,L)$ are locally computable. This means that each $a_k(Q,L)$
can be represented as a sum of volume and boundary integrals of
local invariants constructed from $Q$, $\Omega$, $E$, the curvature
tensor, and their derivatives. Boundary invariants may also include
$S$, $L_{ab}$ and $\chi$. Total mass dimension of such invariants
should be $k$ for the volume terms and $k-1$ for the boundary ones.

At the moment several coefficients of the expansion (\ref{hkexp})
are known for the case of mixed boundary conditions (\ref{mixedbc})
and matrix valued function $Q$ (see \cite{lastpaper} for details of
derivation; the formula (\ref{Qa4bou}) for $a_4$ was derived in
\cite{lastpaper} with additional restrictions $L_{ab}=0$ and $S=0$)
:
\begin{eqnarray}
&&a_0(Q,L)=(4\pi)^{-n/2}\iM \,{\rm tr}\,(Q).\label{Qa0bou} \\
&&a_1(Q,L)={\frac 14}(4\pi)^{-(n-1)/2}\idM
     \,{\rm tr}\, (\chi Q). \label{Qa1bou} \\
&&a_2(Q,L)=\frac 16 (4\pi)^{-n/2}\left\{ \iM
     \,{\rm tr}\,(6QE+QR) \right. \nonumber \\
&&\qquad\qquad \left. +\idM \,{\rm tr} \,
    (2Q L_{aa} +12QS +3 \chi Q_{ ;n})
     \right\} .\label{Qa2bou} \\
&&a_3(Q,L)=\frac 1{384}(4 \pi )^{
       -(n-1)/2}  \idM
       {\rm tr} \big\{ Q( -24 E + 24 \chi E \chi \nonumber\\
&&\qquad\qquad +48 \chi E + 48 E\chi
                     -12 \chi_{ :a} \chi_{:a} + 12 \chi_{:aa}
              -6 \chi_{:a}\chi_{:a}\chi +
              16\chi  R
\nonumber \\ &&\qquad\qquad + 8 \chi R_{anan} +192 S^2 + 96 L_{aa} S
+ (3+10\chi )L_{aa}L_{bb} \nonumber \\ &&\qquad\qquad +(6-4\chi )
L_{ab}L_{ab} ) + Q_{;n}(96 S +192 S^2)
              +24 \chi Q_{ ;nn} \big\}.\label{Qa3bou}
\end{eqnarray}
For a scalar function $Q$ and mixed boundary conditions the
coefficients $a_4$ and $a_5$ were already derived \cite{a4a5}.

\subsection{$\zeta$-function}

 Zeta function of a positive operator $L$ is defined by
\begin{equation}
\zeta_{L}(s) = \sum_\lambda \frac{1}{\lambda^s} \, ,
\end{equation}
where the sum is over all eigenvalues of the operator $L$. The
 zeta function is related to the heat kernel by the transformation
\begin{equation}
\zeta_{L}(s) = \frac{1}{\Gamma(s)} \int_0^{+\infty} dt \, t^{s-1}
K(I, L, t) \, .
\end{equation}
Residues at the poles of the zeta function are related to the
coefficients of the heat kernel expansion:
\begin{equation}
a_k (I, L) = {\rm Res}_{s=(n-k)/2} (\Gamma(s) \zeta_{L}(s) ) \,
.\label{res}
\end{equation}
Here $I$ is a unit matrix with a dimension of the matrix functions
$a^\mu, b$ in (\ref{B4}).From (\ref{res}) it follows that
\begin{equation}
a_n (I, L) = \zeta_{L}(0) \, .
\end{equation}

In  Euclidean four dimensional space the zero temperature one-loop
path integral over the boson fields $\phi = \sum_\lambda C_\lambda
\phi_\lambda$ can be evaluated as follows (up to a normalization
factor):
\begin{equation}
Z = \int d \phi e^{- \int d^4 x \, \phi L\, \phi} \simeq
\prod_\lambda \int \mu d C_\lambda e^{-\lambda C_{\lambda}^2} \simeq
\mu^{\zeta_{L}(0)} {\rm det} L^{-1/2} \, . \label{m1}
\end{equation}
Here we introduced the constant $\mu$ with a dimension of mass in
order to keep a proper dimension of the measure in the functional
integral. $\zeta_L(0)$ can  be thought of as a number of eigenvalues
of the operator $L$. For the operator $L$ in the form (\ref{B4}) the
number of eigenvalues is infinite, so $\zeta_L(0)$ yields a
regularized value for this number.

The zero temperature one-loop effective action is defined then by
\begin{eqnarray}
&& W = - \ln Z = -\frac{1}{2} \ln {\rm det} L + \frac{1}{2}
\zeta_{L}(0) \ln \mu^2 = \frac{1}{2} \zeta_{L}^\prime (0) +
\frac{1}{2}
\zeta_L(0) \ln \mu^2 = \nonumber \\
  && \, \quad = \frac{1}{2}
  \frac{\partial}{\partial s} (\mu^{2s}\zeta_{L}(s))|_{s=0}
\end{eqnarray}

The term $\zeta_L(0) \ln \mu^2 = a_4 (I,L) \ln \mu^2$ in the
effective action $W$ determines the one-loop beta function,  this
term describes renormalization of the one-loop logarithmic
divergences appearing in the theory.

\subsection{Free energy for boson fields}

A finite temperature field theory is defined in Euclidean space,
since for boson fields one has to impose periodic boundary
conditions on imaginary time coordinate (antiperiodic boundary
conditions for fermion fields respectively). A partition function is
defined by
\begin{equation}
Z(\beta) = {\rm Tr}\, e^{-\beta H} \, ,
\end{equation}
where $H$ is a hamiltonian of the problem and $\beta=\hbar/T$. Let
us choose the lagrangian density $\rho$ in the form
\begin{equation}
\rho= -\frac{\partial^2}{\partial \tau^2} + L \, ,
\end{equation}
where $\tau$ is an imaginary time coordinate and $L$ is a three
dimensional spatial part of the density in the form (\ref{B4}). The
free energy of the system is defined by
\begin{equation}
F (\beta) = - \frac{\hbar}{\beta} \ln Z(\beta) =
 - \frac{\hbar}{\beta} \ln \Bigl( N_\beta \int D\phi
 \exp \bigl( -\int_0^\beta d\tau \int d^3 x \,   \phi \rho \phi \bigr) \Bigr)  \,
 ,
\end{equation}
the integration is over all periodic fields satisfying $\phi(\tau +
\beta) = \phi(\tau)$ ($N_\beta$ is a normalization coefficient). As
a result the eigenfunctions of $\rho$ have the form
$\exp(i\tau\omega_n)\phi_\lambda$, where $\omega_n = 2\pi n / \beta$
and $L \phi_\lambda = \lambda \phi_\lambda$. The free energy is thus
equal to \cite{Bordag1}
\begin{equation}
F =  \frac{\hbar}{2\beta} \sum_{n=-\infty}^{+\infty} \sum_\lambda
\ln \frac{(\omega_n^2 + \lambda)}{\mu^2} = - \frac{\hbar}{2\beta}
\frac{\partial}{\partial s} (\mu^{2s}\zeta(s) )|_{s=0} \, ,
\label{m2}
\end{equation}
where we introduced $\zeta$-function
\begin{equation}
\zeta(s) = \sum_{n=-\infty}^{+\infty} \sum_{\lambda} (\omega_n^2 +
\lambda) ^{-s}
\end{equation}
and the parameter $\mu$ with a mass dimensionality in order to make
the argument of the logarithm dimensionless (also see a previous
section).

Then it is convenient to use the formula
\begin{equation}
\zeta(s) = \frac{1}{\Gamma(s)} \int_0^{+\infty} dt \, t^{s-1}
\sum_{n=-\infty}^{+\infty} \sum_{\lambda} e^{-t (\omega_n^2 +
\lambda)} \, ,
\end{equation}
and separate $n=0$ and other terms in the sum. For $n\neq 0$ terms
we substitute the heat kernel expansion for the operator $L$ at
small $t$
\begin{equation}
\sum_{\lambda} e^{-\lambda t} = K(I;L;t) \simeq \sum_{k=0}^\infty
a_k (I,L) t^{(k-3)/2}
\end{equation}
and perform $t$ integration, then we arrive at the high temperature
expansion ($\beta \to 0$) for the free energy $F$:
\begin{eqnarray}
&& F/\hbar = - \frac{1}{2\beta} \zeta_L^\prime(0) -\frac{1}{2\beta}
\zeta_L(0) \ln(\mu^2) + (4\pi)^{3/2} \biggl[ -\frac{a_0}{\beta^4}
\frac{\pi^2}{90} - \frac{a_{1}}{\beta^3}
\frac{\zeta_R(3)}{4\pi^{3/2}} -\frac{a_2}{\beta^2} \frac{1}{24}
\nonumber \\ && + \frac{a_3}{\beta} \frac{1}{(4\pi)^{3/2}}
\ln\Bigl(\frac{\beta\mu} {2\pi}\Bigr)
-\frac{a_4}{16\pi^2}\biggl(\gamma + \ln \frac{\beta\mu}{2\pi}
\biggr) \nonumber \\ && - \sum_{n\ge 5} \frac{a_n}{\beta^{4-n}}
\frac{(2\pi)^{3/2-n}}{2\sqrt{2}} \Gamma \Bigl(\frac{n-3}{2}\Bigr)
\zeta_R (n-3) \biggr] \, . \label{free}
\end{eqnarray}
Here $a_k\equiv a_k(I,L)$, $\zeta_R(s) = \sum_{n=1}^{+\infty}
n^{-s}$ is a Riemann zeta function, $\zeta_L(s)=\sum_{\lambda}
\lambda^{-s}$ is a zeta function of an operator $L$, $\gamma$ is the
Euler constant. The first two terms on the r.h.s. of (\ref{free})
follow from the $n=0$ term.

The term
\begin{equation}
-\frac{(4 \pi)^{3/2} \hbar a_0}{\beta^4} \frac{\pi^2}{90} =
 - V  \: \frac{{\rm tr} I}{ \hbar^3} \frac{\pi^2}{90}  T^4
\end{equation}
is the leading high temperature contribution to the free energy.

The classical limit terms due to the equality $\zeta_L(0) = a_3$ can
be rewritten as follows:
\begin{equation}
T \Bigl( -\frac{1}{2} \zeta_L^\prime(0) + \zeta_L(0) \ln
\frac{\hbar}{2\pi T} \Bigr) = T \, \sum_\lambda \ln \frac{\hbar
\sqrt{\lambda}}{2\pi T} \, . \label{class}
\end{equation}
The terms on the l.h.s. of (\ref{class}) yield a renormalized value
of the terms on the r.h.s. of (\ref{class}), since the sum on the
righthandsight is generally divergent when the number of modes is
infinite.

The term with $a_4$ determines the part of the free energy that
appears due to one-loop logarithmic divergences and thus it depends
on the dimensional parameter $\mu$ as in the zero temperature case.

%The usual definition for the boson free energy can be obtained by
%use of a formula
%\begin{equation}
%\ln(1- e^{-\beta \sqrt{\lambda}}) = -\frac{1}{2}
%\zeta_{\lambda}^\prime (0) - \frac{\beta \sqrt{\lambda}}{2} ,
%\end{equation}
%where
%\begin{equation}
%\zeta_\lambda (s) = \sum_{n=-\infty}^{+\infty} ( \omega_n^2 +
%\lambda )^{-s} \, .
%\end{equation}

\subsection{Chiral anomaly in four dimensions\\
 for MIT bag boundary conditions}

Consider the Dirac operator on an $n$-dimensional Riemannian
manifold
\begin{equation}
  \Dir =\gamma^\mu\left(\partial_\mu+ V_\mu+ iA_\mu\gamma^5- \frac{1}{8} [ \gamma_\rho,
  \gamma_\sigma ] \sigma^{[\rho \sigma]}_\mu \right)
  \label{A2}
\end{equation}
in external vector $V_\mu$ and axial vector $A_\mu$  fields. We
suppose that $V_\mu$ and $A_\mu$ are anti-hermitian matrices in the
space of some representation of the gauge group. $\sigma^{[\rho
\sigma]}_\mu$ is the spin-connection\footnote{The spin-connection
must be included even on a flat manifold if the coordinates are not
Cartesian.}.

The Dirac operator transforms covariantly under infinitesimal local
gauge transformations (the local gauge transformation is $\Dir \to
\exp (-\lambda )\Dir\exp (\lambda )$):
\begin{eqnarray}
&&\delta_\lambda A_\mu =[A_\mu ,\lambda ] \nonumber \\
&&\delta_\lambda V_\mu = \partial_\mu \lambda +[V_\mu ,\lambda ]
\nonumber \\
&&\Dir \to \Dir +[\Dir ,\lambda ] \label{gauge}
\end{eqnarray}
and under infinitesimal local chiral transformations (the local
chiral transformation is $\Dir \to \exp (i\varphi \gamma_5)\Dir\exp
(i\varphi \gamma_5 )$):
\begin{eqnarray}
&&\tilde\delta_\varphi A_\mu =\partial_\mu \varphi +[V_\mu ,\varphi
],
\nonumber \\
&&\tilde\delta_\varphi V_\mu =-[A_\mu ,\varphi ] ,\nonumber \\
&&\Dir \to \Dir + i\{ \Dir , \gamma^5 \varphi \} \,. \label{chiral}
\end{eqnarray}
The parameters $\lambda$ and $\varphi$ are anti-hermitian matrices.

First we adopt the zeta-function regularization and write the
one-loop effective action for Dirac fermions at zero temperature as
\footnote{The one-loop effective action is proportional to Planck
constant $\hbar$, in what following we put $\hbar=1$.}
\begin{equation}
W=-\ln \det \Dir = -\frac 12 \ln \det \Dir^2 = \frac 12
\zeta_{\Dir^{2}}^\prime(0) +\frac 12 \ln (\mu^2) \zeta_{\Dir^{2}}
(0) \,,\label{detD}
\end{equation}
where
\begin{equation}
\zeta_{\Dir^{2}} (s)={\rm Tr} ( \Dir^{-2s} )\,, \label{zeta}
\end{equation}
prime denotes differentiation with respect to $s$, and ${\rm Tr}$ is
the functional trace.

The following identity holds:
\begin{equation}
\zeta_{A}(s)= {\rm Tr} A^{-s} \Rightarrow \delta \zeta_{A}(s)=-s
{\rm Tr}((\delta A)A^{-s-1})\, . \label{zeta1}
\end{equation}

Due to the identity (\ref{zeta1})
\begin{equation}
\delta_\lambda \zeta_{{\Dir^{2}}}(s) =- \left( 2s {\rm Tr} (
[\Dir,\lambda ] \Dir^{-2s-1}) \right) = -2s \left( {\rm Tr} (
[\Dir^{-2s},\lambda ]) \right) =0\, ,
\end{equation}
so the effective action (\ref{detD}) is gauge invariant,
$\delta_\lambda W=0$.

The chiral anomaly is by definition equal to the variation of $W$
under an infinitesimal chiral transformation. Using (\ref{zeta1}) we
obtain:
\begin{equation}
\tilde\delta_\varphi \zeta_{{\Dir^{2}}}(s) = -\left( 2is {\rm Tr} (
\{\Dir, \gamma^5\varphi\} \Dir^{-2s-1}) \right) = -4is \left( {\rm
Tr} (
 \gamma^5 \varphi \Dir^{-2s}) \right) \, ,
\end{equation}
and the anomaly reads
\begin{equation}
\mathcal{A}:=\tilde\delta_\varphi W = \frac 12
\tilde\delta_\varphi\zeta_{\Dir^{2}}^\prime (0)= -2 {\rm Tr} (i
\gamma^5 \varphi \Dir^{-2s} )\vert_{s=0} \,. \label{chian}
\end{equation}

The heat kernel is related to the zeta function by the Mellin
transformation:
\begin{equation}
{\rm Tr} (i \gamma^5 \varphi \Dir^{-2s} )=\Gamma (s)^{-1}
\int_0^\infty dt\, t^{s-1} K(i \gamma^5\varphi, \Dir^2, t) \,.
\label{zeta2}
\end{equation}
In particular, after the substitution of the heat kernel expansion
(\ref{hkexp}) into the formula (\ref{zeta2}) we obtain
\begin{equation}
\mathcal{A} =-2 a_n (i \gamma^5\varphi,\Dir^2) \,.\label{anomhk}
\end{equation}
The same expression for the anomaly follows also from the Fujikawa
approach \cite{Fujikawa:1979ay}.

One can also derive the expression for the anomaly (\ref{anomhk})
from Schwinger's effective action.  One should start from an
identity:
\begin{equation}
\ln \lambda = - \int_{0}^{+\infty} \frac{dt}{t} e^{-t\lambda}
\end{equation}
Then the change in the effective action due to chiral
transformations can be written:
\begin{align}
\mathcal{A} &= \tilde\delta_\varphi \Bigl(-\frac{1}{2}\ln\det\Dir^2
\Bigr) = \tilde\delta_\varphi {\rm tr} \int_{0}^{+\infty}
\frac{dt}{2t} \,\,\, e^{-t\Dir^2} = \nonumber \\ &= -i {\rm tr}
\int_{0}^{+\infty} \{ \gamma^5 \varphi , \Dir^2 \} e^{-t\Dir^2} = 2i
{\rm tr} \gamma^5 \varphi \int_{0}^{+\infty}
\frac{\partial}{\partial t} e^{-t\Dir^2} = \nonumber \\
&= -2 \lim_{t\to 0} {\rm tr} i \gamma^5 \varphi \; e^{-t\Dir^2} = -2
a_n (i \gamma^5\varphi,\Dir^2) .
\end{align}

We impose local boundary conditions:
\begin{equation}
\Pi_-\psi \BB =0,\qquad \Pi_-=\frac 12 \left( 1-\gamma^5 \gamma_n
\right) \,,\label{bagbc}
\end{equation}
which are nothing else than a Euclidean version of the MIT bag
boundary conditions \cite{bag}. For these boundary conditions
$\Pi_-^\dag =\Pi_-$, and the normal component of the fermion current
$\psi^\dag \gamma_n \psi$ vanishes on the boundary. Spectral
properties of the Dirac operator for bag boundary conditions are
intensively studied \cite{Santangelo}.

Since $\Dir$ is a first order differential operator it was enough to
fix the boundary conditions (\ref{bagbc}) on a half of the
components. To proceed with a second order operator $L=\Dir^2$ we
need boundary conditions on the remaining components as well. They
are defined by the consistency condition \cite{BG-Dirac}:
\begin{equation}
\Pi_-\Dir \psi \BB =0 \,,\label{conscond}
\end{equation}
which is equivalent to the Robin boundary condition
\begin{equation}
 \left(\nabla_n + S\right) \Pi_+ \psi \BB =0 \, ,\,\qquad \Pi_+=\frac 12 \left( 1+\gamma^5 \gamma_n
\right)
\end{equation}
 with
\begin{equation}
S= - \frac{1}{2}\Pi_+L_{aa} \,.\label{B11}
\end{equation}

In the paper \cite{lastpaper} the following expression for a
coefficient $a_4 (Q, L)$  with an hermitian matrix valued function
$Q$ and conditions (\ref{mixedbc}), $L_{ab}=0$ (flat boundaries),
$S=0$ was obtained:
\begin{eqnarray}
&&a_4(Q,L)=\frac 1{360} (4 \pi )^{
       -n/2} \Big\{ \iM \,{\rm tr}\,
       \big\{ Q(60{E_{ ; \mu}}^\mu+60 R E+180E^2 \nonumber \\
&&\qquad\qquad +30 \Omega_{\mu\nu}
      \Omega^{\mu\nu} +12 {R_{;\mu}}^\mu +
      5 R^2-2R_{\mu\nu}R^{\mu\nu}+2R_{\mu\nu\rho\sigma}R^{\mu\nu\rho\sigma})
\big\} \nonumber \\
&&\qquad\qquad+ \idM \,{\rm tr} \,
      \big\{ Q \{ 30 E_{;n} + 30 \chi E_{;n} \chi +
      90 \chi E_{;n} + 90 E_{;n}\chi  \nonumber\\
&&\qquad\qquad +18\chi\chi_{:a}\Omega_{an} + 12
\chi_{:a}\Omega_{an}\chi +
  18 \Omega_{an}\chi\chi_{:a}  -  12 \chi\Omega_{an}\chi_{:a}
  \nonumber\\ &&\qquad\qquad
         + 6 [\chi\Omega_{an}\chi, \chi_{:a}] + 54
[\chi_{:a}, \Omega_{an}] + 30 [\chi , \Omega_{an:a}] + 12 R_{ ;n} +
30 \chi R_{ ;n} \} +
 \nonumber\\ &&\qquad\qquad
       + Q_{ ;n}(- 30 E + 30 \chi E \chi + 90 \chi E
+ 90 E \chi - \nonumber \\
&&\qquad\qquad
 -18 \chi_{ :a} \chi_{ :a}   + 30 \chi_{:aa}
 - 6 \chi_{:a}\chi_{:a}\chi + 30
\chi R )+  30 \chi {Q_{ ;\mu}}^{\mu n}  \big\} \Big\} .
\label{Qa4bou}
\end{eqnarray}

To obtain the chiral anomaly in four dimensions\footnote{In two
dimensions ($n=2$) the boundary part of the chiral anomaly with MIT
bag boundary conditions is equal to zero \cite{lastpaper}.} with MIT
bag boundary conditions one has to calculate the coefficient $a_4(Q,
L)$ (\ref{Qa4bou}) with $L=\Dir^2$, $Q= i\gamma^5 \phi$ and
substitute it into (\ref{anomhk}). We define $V_{\mu\nu} =
\partial_{\mu} V_{\nu} -
\partial_{\nu} V_{\mu} + [V_{\mu}, V_{\nu}]$,
$A_{\mu\nu} = D_{\mu} A_{\nu} - D_{\nu} A_{\mu}$, $D_{\mu} A_{\nu} =
\partial_{\mu} A_{\nu} - \Gamma^{\rho}_{\mu\nu}A_{\rho}
+ [V_{\mu},A_{\nu}]$. The anomaly contains two contributions:
\begin{equation}
\mathcal{A}=\mathcal{A}_V + \mathcal{A}_b \,. \label{Ain4}
\end{equation}
In the volume part
\begin{eqnarray}
    \label{B16}
&&\mathcal{A}_V=
    \frac{-1}{180 \, (2 \pi)^{2} } \int_M \, d^{4}x \sqrt{g} \,
    {\rm tr} \varphi
     \Bigl( - 120 \, [D_{\mu}V^{\mu\nu},A_\nu] \nonumber\\
&&\qquad\qquad  +60 \, [D_{\mu}A_\nu,V^{\mu\nu}]
   - 60 \, D_{\mu}D^{\mu}D_{\nu}A^\nu
   + 120 \, \{\{D_{\mu}A_\nu , A^\nu\} , A^\mu\}\nonumber\\
&&\qquad\qquad   + 60 \, \{D_{\mu} A^\mu,A_\nu A^\nu \}
    + 120 \, A_\mu D_{\nu}A^\nu A^\mu
   + 30 \, [[A_\mu , A_\nu],A^{\mu\nu}]   \nonumber \\
&&\qquad\qquad    + \epsilon_{\mu\nu\rho\sigma}\
   \{ - 45 \, i\, V^{\mu\nu}V^{\rho\sigma}
    + 15 \, i \, A^{\mu\nu}A^{\rho\sigma}
   - 30 \, i \, (V^{\mu\nu}A^{\rho} A^{\sigma}
   + A^{\mu} A^{\nu} V^{\rho\sigma})
   \nonumber  \\
&&\qquad\qquad   - 120 \, i \, A^{\mu} V^{\nu\rho} A^{\sigma}
     + 60 \, i \, A^{\mu} A^{\nu} A^{\rho} A^{\sigma}\}
   - 60 \, (D_{\sigma}A_\nu) R^{\nu\sigma}
  + 30 \, (D_{\mu}A^\mu) R \nonumber \\
&&\qquad\qquad
   - \frac{15i}{8} \epsilon_{\mu\nu\rho\sigma}
\, R^{\mu\nu}{}_{\eta \theta}
    R^{\rho\sigma \eta \theta}
    \Bigr )
\end{eqnarray}
only the $DA - R$ terms seem to be new \cite{lastpaper} (for flat
space it can be found e.g. in \cite{Andrianov:1983fg}).

The boundary part
\begin{eqnarray}
&&\mathcal{A}_b=
    \frac{-1}{180 \, (2\pi)^2} \int_{\partial M} \, d^{3}x \sqrt{h}
    \,   {\rm tr} \Bigl(
    12 \, i \, \epsilon^{abc} \, \{A_b, \varphi \} D_a A_c
\nonumber \\
 &&\qquad\qquad   + 24 \{\varphi, A^a \} \{A_a, A_n\}
    -60 \, [A^a, \varphi] (V_{na} - [A_n, A_a]) \nonumber\\
&&\qquad\qquad  + 60 (D_n \varphi ) D_{\mu} A^{\mu}  \Bigr)
\label{banomaly}
   \end{eqnarray}
is new \cite{lastpaper}. It has been derived under the two
restrictions: $S=0$ and $L_{ab}=0$. Note, that in the present
context, the first condition ($S=0$) actually follows from the
second one ($L_{ab}=0$) due to (\ref{B11}).

\section{Casimir effect for rectangular cavities}
\subsection{Casimir energy of two perfectly conducting
parallel plates}

The Casimir energy is usually defined as
\begin{equation}
E=\sum_i \frac{\hbar \omega_i}{2} , \label{E1}
\end{equation}
where the sum is over all eigenfrequencies of the system. In what
following we put $\hbar=1$.
 We start
from the well known case of two perfectly conducting plates
separated by a distance $a$ from each other. In this case the
eigenfrequencies $\omega_i$ are defined as follows:
\begin{eqnarray}
&& \omega_{TE} = \sqrt{(\pi n/a)^2+ k_x^2+ k_y^2} , \, n=1 ..
+\infty \\
&& \omega_{TM} = \sqrt{(\pi n/a)^2+ k_x^2+ k_y^2} , \, n=1 ..
+\infty \\
&& \omega_{\rm main wave} = \sqrt{k_x^2 + k_y^2} ,
\end{eqnarray}
so that the Casimir energy can be written as
\begin{equation}
E = \frac{S}{2} \Biggl(\sum_{n=1}^{+\infty} +
\sum_{n=0}^{+\infty}\Biggr) \iint_{-\infty}^{+\infty} \frac{dk_x
dk_y}{(2\pi)^2} \sqrt{(\pi n/a)^2+ k_x^2+ k_y^2} , \label{f2}
\end{equation}
$S$ is the surface of each plate. The first sum is equivalent to the
sum over eigenfrequencies of the scalar field satisfying Dirichlet
boundary conditions, the second sum is equivalent to the sum over
eigenfrequencies of the scalar field satisfying Neumann boundary
conditions.

 The expression for the Casimir energy written in this
form is divergent. One has to regularize it somehow to obtain a
finite answer for the energy. Different methods were used for this
purpose. In the present paper we suggest a method which makes
calculations of determinants straightforward and easy to perform.

By making use of an identity
\begin{equation}
\int_{-\infty}^{+\infty} \frac{ds}{2\pi} \ln \frac{s^2+k^2}{s^2} = k
\end{equation}
we can see that up to an irrelevant constant the Casimir energy can
be written in the form (we introduce a dimensional parameter $\mu$
by the same reasoning as in (\ref{m1}) or (\ref{m2})):
\begin{equation}
\begin{split}
E &= \frac{S}{2} \sum_{n=-\infty}^{+\infty}
\iiint_{-\infty}^{+\infty} \frac{dk_x dk_y ds}{(2\pi)^3} \ln
\Biggl(\frac{(\pi n/a)^2+ k_x^2+ k_y^2 + s^2}{\mu^2} \Biggr)= \\
&= \frac{S}{(2\pi)^2} \Biggl(\sum_{n=1}^{+\infty} +
\sum_{n=0}^{+\infty}\Biggr) \int_{0}^{+\infty} dk \,\,\, k^2 \,
\ln \Bigl( n^2+ (ka/\pi)^2 \Bigr) - \\ & \quad -\frac{1}{T}\ln(a
\mu /\pi) a_4(I,L). \label{f3}
\end{split}
\end{equation}
Now the expression for the Casimir energy is written in the standard
$\rm Tr ln = ln Det$ form, which is usual for one-loop effective
actions in quantum field theory. The coefficient $a_4(I,L)$ is equal
to zero for our current choice of the operator $L$ and boundary
geometry.

At this point we introduce a regularization - we restrict
integrations over momenta by some cut off $K$ in the momentum space.
The sums over $n$ are also restricted as follows:
\begin{equation}
\sum_{n=1}^{+\infty} + \sum_{n=0}^{+\infty} \rightarrow
\sum_{n=1}^{+N} + \sum_{n=0}^{+N} .
\end{equation}
The regularized Casimir energy is defined by:
\begin{equation}
E_{reg} = \frac{S}{(2\pi)^2} \Biggl(\sum_{n=1}^{+N} +
\sum_{n=0}^{+N}\Biggr) \int_{0}^{+K} dk \,\,\, k^2 \, \ln \Bigl(
n^2+ (ka/\pi)^2 \Bigr) . \label{r4}
\end{equation}

It is convenient to perform a summation over $n$ first. The
following identity holds:
\begin{equation}
\sum_{n=1}^{N} \ln (n^2 + (ka/\pi)^2) = \sum_{n=1}^{N} \ln (1+
(ka)^2/\pi^2 n^2) + \sum_{n=1}^{N} \ln ( n^2 ) . \label{r2}
\end{equation}
The first sum in (\ref{r2}) can be calculated in the $N \to
+\infty$ limit by use of an identity:
\begin{equation}
\prod_{n=1}^{+\infty} \Bigl(1+ (k a)^2/\pi^2 n^2 \Bigr) =
\frac{\sinh (ka)}{ka} .
\end{equation}
The second sum in (\ref{r2}) can be derived by use of a Stirling
formula (which is exact in the large $N$ limit):
\begin{equation}
N ! \sim \sqrt{2\pi} N^{N+1/2} e^{-N} ,
\end{equation}
so in the large $N$ limit it is possible to write:
\begin{equation}
\sum_{n=1}^{N} \ln ( n^2 ) = 2 \ln N! = \ln (2\pi) + f(N).
\end{equation}
In the large $N$ limit the Dirichlet sum (\ref{r2}) can be
rewritten as:
\begin{equation}
\sum_{n=1}^{N} \ln (n^2 + (ka/\pi)^2) = ka + \ln (1-\exp(-2ka)) -
\ln(k a/\pi) + f(N) . \label{r5}
 \end{equation}
The sum over Neumann modes can be rewritten as follows:
\begin{equation}
\sum_{n=0}^{N} \ln (n^2 + (ka/\pi)^2) = ka + \ln (1-\exp(-2ka)) +
\ln(k a/\pi) + f(N) . \label{r6}
\end{equation}

It is possible to add any finite number that does not depend on
$a$ to the regularized Casimir energy $E_{reg}$ (\ref{r4}) (the
force between the plates is being measured in experiments, so the
energy can be defined up to a constant). We add the surface term
\begin{equation}
 - \frac{S}{(2\pi)^2}\int_{0}^{K} d k \,\,\, k^2
   \:\:\: 2 f(N)  \label{surf1}
\end{equation}
to the regularized Casimir energy $E_{reg}$ (\ref{r4}). Doing so
we obtain
\begin{equation}
E_{reg} = 2 \frac{S a}{(2\pi)^2} \int_{0}^{K} dk \,\, k^3 +
\frac{S}{(2\pi)^2} \int_{0}^{K} d k \,\, k^2 \,\, 2 \ln \Bigl(1 -
\exp(- 2 k a) \Bigr)  \label{EV}
\end{equation}
The first term in (\ref{EV}) is twice the regularized Casimir
energy of the free scalar field since it can be rewritten as
\begin{equation}
2 \frac{V}{(2\pi)^3} \int_{0}^{K} d k \,\, 4\pi k^2 \,\,\,
\frac{k}{2} . \label{EV2}
\end{equation}
This term should be subtracted because we are interested in the
change of the ground state energy when the plates are inserted into
the free space.

Next we perform the limits $K\to +\infty, N\to +\infty$. The
Casimir energy is thus
\begin{equation}
E=\frac{S}{(2\pi)^2} \int_{0}^{+\infty} d k \,\, k^2 \,\, 2 \ln
\Bigl(1 - \exp(- 2 k a) \Bigr) = - \frac{S \pi^2}{720 a^3} ,
\label{plates}
\end{equation}
which is the well known result by Casimir \cite{Casimir}.

After elaborations we summarize the key points of the method,
which is valid for the calculations in cylindrical cavities with
arbitrary cross sections. Suppose we want to calculate $\rm Trln$
of the second order operator $L^{(4)} = L^{(1)} + L^{(3)}$, where
the dimensionalities of the operators are denoted by numbers. At
zero temperature in our case of interest the operator $L^{(3)}$
describes a scalar field inside an infinite waveguide of an
arbitrary cross section with Dirichlet or Neumann boundary
conditions imposed. The eigenmodes of the operators $L^{(1)}$ and
$L^{(3)}$ are denoted by $\lambda_i^{(1)}$ and $\lambda_k^{(3)}$
respectively. The following expression is finite (as can be seen
from the heat kernel expansion):
\begin{equation}
 \frac{1}{2} \ln \prod_{i} \frac{\lambda_i^{(1)} +
  \lambda_k^{(3)}}{\lambda_i^{(1)}} \equiv
\frac{1}{2} \ln \prod_{i} \Biggl(1 +
  \frac{\lambda_k^{(3)}}{\lambda_i^{(1)}} \Biggr) . \label{f4}
\end{equation}
To obtain the initial determinant one should add to (\ref{f4}) the
term
\begin{equation}
\frac{1}{2} \ln \prod_{i}^{N} \lambda_i^{(1)} = s(N) + const
\label{r1}
\end{equation}

 The sum of (\ref{f4}) and (\ref{r1}) generally
 has the following structure (to obtain the total Casimir energy
 the sum over indices $k$ and $p$ has to be performed):
\begin{equation}
\frac{a\sqrt{ \lambda_k^{(3)}}}{2} + g \bigl(\lambda_k^{(3)} a^2
\bigr) + h^{surf}_{Dir\: } (\lambda_p^{(2)}/\mu^2) +
h^{surf}_{Neum\:} (\lambda_p^{(2)}/\mu^2) + 2 s(N), \label{f1}
\end{equation}
where
\begin{equation}
\sum_k g \bigl(\lambda_k^{(3)} a^2 \bigr) - {\rm a \,\,\, convergent
\,\,\, sum,}  \label{f5}
\end{equation}
the term (\ref{f5}) yields the energy of interaction for two flat
parallel plates separated by a distance $a$ inside an infinite
waveguide of the same cross section as these parallel plates (the
walls of a perfectly conducting waveguide are perpendicular to two
flat parallel perfectly conducting plates inside it). The term
(\ref{f5}) yields an experimentally measurable contribution to the
Casimir energy of the cavity (see Sec. $3.7$ for details).

The  term $\sum_k a\sqrt{\lambda_k^{(3)}}/2$ is equal to the
self-energy of an infinite waveguide when $a \to \infty$. For
rectangular cavities the term $\sum_k a\sqrt{\lambda_k^{(3)}}/2$
can be transformed to ${\rm Trln} L_2^{(4)}$ in the same manner as
in the beginning of this section (see a transition from (\ref{f2})
to (\ref{f3}) ). For the operator $L_2^{(4)}$ we repeat the step
(\ref{f4}) and continue this cycle until the first term in the
righthandsight of (\ref{f1}) gets the form of the vacuum energy in
an infinite space, i.e. the form (\ref{EV2}).

The $h^{surf}$ terms describe the self-energies of two parallel
plates inside the waveguide due to Dirichlet and Neumann modes,
these self-energies do not depend on $a$. For flat boundaries
Dirichlet and Neumann boundary contributions to the Casimir energy
cancel each other identically as can be seen from the Seeley
coefficient $a_1$ (\ref{Qa1bou}), for two parallel plates it can
be seen from the expressions (\ref{r5}) and (\ref{r6}) .

A contribution from the last term $s(N)$ is proportional to $a_3
(I, L^{(3)} )$ (at zero temperature it is  just the effective
number of modes inside an infinite perfectly conducting waveguide,
and thus it is not relevant to the energy of interaction between
the two plates inside the waveguide) and $a_4(I, L^{(4)})$ (this
term is also not relevant when the interaction of the two plates
inside an infinite waveguide is studied).

To implement (\ref{f4}) we used the following equality:
\begin{equation}
\prod_{n=1}^{+\infty} \frac{(\pi n/a)^2 + \lambda_k^{(3)}}{(\pi
n/a)^2} \equiv \prod_{n=1}^{+\infty} \Bigl(1 +
\frac{\lambda_k^{(3)}a^2}{\pi^2 n^2}\Bigr) = \frac{\sinh a
\sqrt{\lambda_k^{(3)}}}{a \sqrt{\lambda_k^{(3)}}} . \label{f6}
\end{equation}

\subsection{Casimir energy of a perfectly conducting
rectangular waveguide} For a perfectly conducting rectangular
waveguide the technical issues can be done in analogy with two
parallel plates. We tacitly assume that the reader understood how
the regularization is introduced in our method, so we will write
only main steps without bothering too much on divergent form of
some expressions. The Casimir energy for unit length is:
\begin{equation}
E = \frac{1}{2} \sum_{n_1} \sum_{n_2} \iint_{-\infty}^{+\infty}
\frac{d p_1 d p_2}{(2\pi)^2} \ln \Bigl( (\pi n_1/a)^2 + (\pi
n_2/b)^2 + p_1^2 + p_2^2 \Bigr)
\end{equation}
For {\rm TM} modes $n_1$ and $n_2$ take positive integer values from
$1$ to $+\infty$, for {\rm TE}  modes $n_1$ and $n_2$ take positive
integer values and one of them can be equal to zero ($n_1=n_2=0$
corresponds to the main wave case).

So the energy can be rewritten as:
\begin{align}
E &= \frac{1}{4} \sum_{n_1, n_2 = -\infty}^{+\infty} \iint \frac{d
p_1 d p_2}{(2\pi)^2} \ln \Bigl( (\pi n_1/a)^2 + (\pi n_2/b)^2 +
p_1^2 +
p_2^2 \Bigr) = \nonumber \\
&= \frac{1}{4 a b} \sum_{n_1, n_2 =-\infty}^{+\infty} \iint \frac{d
p_1 d p_2}{(2\pi)^2} \ln \Bigl( (\pi n_1)^2 t + (\pi n_2)^2
\frac{1}{t} + p_1^2 + p_2^2 \Bigr) ,
\end{align}
where $t=b/a$ or $t=a/b$. Using formula (\ref{f6}) for
multiplication over $n_1$ we obtain for the energy:
\begin{align}
E &= \frac{1}{4 a b} \sum_{n_2=-\infty}^{+\infty} \iint \frac{d p_1
d p_2}{(2\pi)^2} \,\,\, 2 \ln\sinh \sqrt{(\pi n_2/t)^2 +
\frac{p_1^2+p_2^2}{t} }  =  \\
 &= \frac{1}{4 a b} \sum_{n_2=-\infty}^{+\infty} \iint \frac{d p_1
d p_2}{(2\pi)^2} \,\,\,\Biggl[ 2 \sqrt{(\pi n_2/t)^2 +
\frac{p_1^2+p_2^2}{t} } - (2\ln 2) + \label{f7}\\
&+   2 \ln \biggl(1- \exp\Bigl(-2 \sqrt{(\pi n_2/t)^2 +
\frac{p_1^2+p_2^2}{t} }\Bigr)\biggr)\Biggr] . \label{f8}
\end{align}
A contribution from the term $(-2\ln2)$  in (\ref{f7}) should be
subtracted following the analysis of Section $3.1$. The part
(\ref{f8}) with the logarithm is finite, it contributes to the
finite final answer for the Casimir energy.

For the first term in (\ref{f7}) we get:
\begin{align}
& \frac{1}{2 a b} \sum_{n_2=-\infty}^{+\infty} \iint \frac{d p_1 d
p_2}{(2\pi)^2} \,\,\,  \sqrt{(\pi n_2/t)^2 + \frac{p_1^2+p_2^2}{t} }
=  \\ &= \frac{1}{2 t^2 ab} \sum_{n_2=-\infty}^{+\infty} \iiint
\frac{d p_1 d p_2 d p_3}{(2\pi)^3} \,\,\, \ln \Bigl((\pi n_2)^2 +
p_1^2 + p_2^2 + p_3^2 \Bigr) = \label{f9} \\
&= -\frac{\pi^2}{720 t^2 ab} \label{f10}
\end{align}
because up to a numerical coefficient the expression (\ref{f9}) is
just the same as the formula (\ref{f3}).

So the Casimir energy for unit length of a rectangular waveguide can
be written as the sum of (\ref{f8}) and (\ref{f10}) :
\begin{align}
 &E_{waveguide} (a, b) = \nonumber \\ &= -\frac{\pi^2}{720 t^2 ab}
+ \frac{t}{4\pi ab}
 \sum_{n=-\infty}^{+\infty} \int_{0}^{+\infty} dp \,\, p \,\,
 \ln \biggl( 1 - \exp\Bigl(-2\sqrt{\frac{\pi^2 n^2}{t^2} + p^2 }
 \Bigr) \biggr) \label{wave}
\end{align}

\subsection{Casimir energy of a perfectly conducting
rectangular cavity}

The Casimir energy in this case can be written as:
\begin{equation}
\begin{split}
E &= \sum_{n_1, n_2, n_3 = 1}^{+\infty} \sqrt{\Bigl(\frac{\pi
n_1}{a}\Bigr)^2 + \Bigl(\frac{\pi n_2}{b}\Bigr)^2 + \Bigl(\frac{\pi
n_3}{c}\Bigr)^2} + \\ &\quad + \frac{1}{2} \sum_{n_1, n_2 =
1}^{+\infty} \sqrt{\Bigl(\frac{\pi n_1}{a}\Bigr)^2 + \Bigl(\frac{\pi
n_2}{b}\Bigr)^2} + \\ &\quad + \frac{1}{2} \sum_{n_1, n_3 =
1}^{+\infty} \sqrt{\Bigl(\frac{\pi n_1}{a}\Bigr)^2 + \Bigl(\frac{\pi
n_3}{c}\Bigr)^2} + \\ &\quad +\frac{1}{2} \sum_{n_2, n_3 =
1}^{+\infty} \sqrt{\Bigl(\frac{\pi n_2}{b}\Bigr)^2 + \Bigl(\frac{\pi
n_3}{c}\Bigr)^2} = \\
&= \frac{1}{8} \sum_{n_1, n_2, n_3 = -\infty}^{+\infty}
\sqrt{\Bigl(\frac{\pi n_1}{a}\Bigr)^2 + \Bigl(\frac{\pi
n_2}{b}\Bigr)^2 + \Bigl(\frac{\pi n_3}{c}\Bigr)^2} - \\
&\quad - \sum_{n_1=-\infty}^{+\infty} \frac{1}{8}
\sqrt{\Bigl(\frac{\pi n_1}{a}\Bigr)^2} -
\sum_{n_2=-\infty}^{+\infty} \frac{1}{8} \sqrt{\Bigl(\frac{\pi
n_2}{b}\Bigr)^2} - \sum_{n_3=-\infty}^{+\infty} \frac{1}{8}
\sqrt{\Bigl(\frac{\pi n_3}{c}\Bigr)^2} \label{f22}
\end{split}
\end{equation}
Using formula (\ref{f6}) and technique described in previous
subsections we obtain:
\begin{equation}
\begin{split}
&\frac{1}{8} \sum_{n_1, n_2, n_3 = -\infty}^{+\infty}
\sqrt{\Bigl(\frac{\pi n_1}{a}\Bigr)^2 + \Bigl(\frac{\pi
n_2}{b}\Bigr)^2 + \Bigl(\frac{\pi n_3}{c}\Bigr)^2} = a E_{waveguide}
(b,c) +
\\ &+ \frac{1}{4} \sum_{n_2, n_3 = -\infty}^{+\infty}
\int_{-\infty}^{+\infty} \frac{dp}{2\pi} \ln \biggl(1 - \exp
\Bigl(-2 a \sqrt{\bigl(\frac{\pi n_2}{b}\bigr)^2 + \bigl(\frac{\pi
n_3}{c}\bigr)^2 + p^2} \Bigr)\biggr)
\end{split}
\end{equation}

The remaining terms should be calculated (using formula (\ref{f6})
again) as follows:
\begin{align}
\sum_{n_1=-\infty}^{+\infty} \frac{1}{8} \sqrt{\Bigl(\frac{\pi
n_1}{a}\Bigr)^2} &= \frac{1}{8} \sum_{n_1=-\infty}^{+\infty}
 \int_{-\infty}^{+\infty} \frac{dp}{2\pi} \ln \biggl(\Bigl(\frac{\pi
n_1}{a}\Bigr)^2 + p^2 \biggr) = \nonumber \\
&= \frac{1}{4} \int_{-\infty}^{+\infty} \frac{dp}{2\pi} \ln \Bigl(1-
\exp(-2 a p)\Bigr) = - \frac{\pi}{48 a} \label{f21}
\end{align}

As a result for the Casimir energy of the cavity we obtain:
\begin{multline}
\begin{split}
&E_{cavity} (a,b,c) =  \frac{\pi}{48 a} + \frac{\pi}{48 b} +
\frac{\pi}{48 c} + a E_{waveguide}(b,c) + \\ &+ \frac{1}{4}
\sum_{n_2, n_3 = -\infty}^{+\infty} \int_{-\infty}^{+\infty}
\frac{dp}{2\pi} \ln \biggl(1 - \exp \Bigl[-2 a \sqrt{\Bigl(\frac{\pi
n_2}{b}\Bigr)^2 + \Bigl(\frac{\pi n_3}{c}\Bigr)^2 + p^2}
\Bigr]\biggr) . \label{box}
\end{split}
\end{multline}

\subsection{Relation to the argument principle}
An argument principle is a convenient method of summation over the
eigenmodes of the system (see \cite{Ginzburg} and \cite{Elizalde}
for its applications). The argument principle states:
\begin{equation}
\frac{1}{2\pi i} \oint \phi(\omega) \frac{d}{d\omega} \ln f(\omega)
d\omega = \sum \phi (\omega_0) -\sum \phi(\omega_\infty) ,
\end{equation}
where $\omega_0$ are zeroes and $\omega_\infty$ are poles of the
function $f(\omega)$ inside the contour of integration. For the
Casimir energy $\phi(\omega) = \omega/2$. We choose
\begin{equation}
f(\omega) = \frac{2 \sin \biggl[ a \sqrt{\omega^2 - k_x^2 -
k_y^2}\biggr] }{\mu \sqrt{\omega^2 - k_x^2 - k_y^2}}
\end{equation}
in case of a scalar field satisfying Dirichlet boundary conditions
on the plates. The contour lies on an imaginary axis, a contribution
from the right semicircle with a large radius is negligible. A
denominator is chosen in this form to remove $\omega^2=k_x^2+k_y^2$
from the roots of the equation $f(\omega)=0$. In this case we
proceed as follows:
\begin{equation}
\begin{split}
E_{Dir} &= \frac{-S}{2\pi i} \iint \frac{dk_x dk_y}{(2\pi)^2}
\int_{-i \infty}^{+i \infty} d\omega \frac{\omega}{2}
\frac{\partial}{\partial \omega} \ln \frac{2 \sin \biggl[ a
\sqrt{\omega^2 - k_x^2 - k_y^2}\biggr]}
{\mu \sqrt{\omega^2 - k_x^2 - k_y^2}  } = \\
&= \frac{S}{4\pi i} \iint \frac{dk_x dk_y}{(2\pi)^2} \int_{-i
\infty}^{+i \infty} d\omega \ln \frac{ 2 \sin \biggl[ a
\sqrt{\omega^2 - k_x^2 - k_y^2}\biggr]}{\mu \sqrt{\omega^2 - k_x^2 - k_y^2}  } = \\
&= \frac{S}{2} \iiint \frac{d k_x d k_y d\Tilde{\omega}}{(2\pi)^3}
\ln \frac{2 \sinh\biggl[ a \sqrt{\Tilde{\omega}^2 + k_x^2 + k_y^2}
\biggr]}{\mu \sqrt{\Tilde{\omega}^2 + k_x^2 + k_y^2} } = \\
&= \frac{S}{(2\pi)^2} \int_{0}^{+\infty} dk \; k^2 \ln\Bigl(1-
\exp(-2ka) \Bigr) + \\ &\quad +\frac{V}{(2\pi)^3} \int_{0}^{K} d k
\,\, 4\pi k^2 \;  \frac{k}{2}
- \frac{S}{(2\pi)^2} \int_{0}^{K} dk \; k^2 \ln(\mu k)  = \\
&= - \frac{S\pi^2}{1440 a^3} + \text{volume free space contribution}
+ \\ &\quad +\text{surface contribution} .
\end{split}
\end{equation}
Here $\omega= i \Tilde{\omega}$. We see that the argument
principle is in agreement with (\ref{f6}).

\subsection{Zeta function regularization for the cavity}
$\zeta$-function has already been discussed in this paper, so it is
natural to describe regularization of the Casimir energy for the
cavity in terms of $\zeta$-function. Usually the Casimir energy is
regularized as follows:
\begin{equation}
E = \frac{1}{2} \sum_{\omega_l}  \omega_l^{-s} , \label{f23}
\end{equation}
where $s$ is large enough to make (\ref{f23}) convergent. Then we
should continue analytically (\ref{f23}) to the value $s=-1$ , this
procedure yields the renormalized finite Casimir energy. In our case
eigenfrequencies $\omega_l$ should be taken from (\ref{f22}). So the
regularized Casimir energy of the cavity $E_{cavity}(a,b,c,s)$ can
be written in terms of Epstein $Z_3 \bigl(\frac{1}{a}, \frac{1}{b},
\frac{1}{c} ; s \bigr)$ and Riemann $\zeta_R(s)$  zeta functions:
\begin{equation}
\begin{split}
E_{cavity}(a,b,c,s) &= \frac{\pi}{8} \Biggl(
\sum_{n_1,n_2,n_3=-\infty}^{+\infty\:\:\prime}
\Bigl[\Bigl(\frac{n_1}{a}\Bigr)^2 + \Bigl(\frac{n_2}{b}\Bigr)^2 +
\Bigl(\frac{n_3}{c}\Bigr)^2 \Bigr]^{-s/2} - \\
& \quad - 2 \Bigl(\frac{1}{a} + \frac{1}{b} + \frac{1}{c} \Bigr)
\sum_{n=1}^{+\infty} \frac{1}{n^s} \Biggr)
\end{split}
\end{equation}
\begin{align}
 Z_3 \Bigl(\frac{1}{a},
\frac{1}{b}, \frac{1}{c} ; s \Bigr) &=
\sum_{n_1,n_2,n_3=-\infty}^{+\infty\:\:\prime}
\Bigl[\Bigl(\frac{n_1}{a}\Bigr)^2 + \Bigl(\frac{n_2}{b}\Bigr)^2 +
\Bigl(\frac{n_3}{c}\Bigr)^2 \Bigr]^{-s/2} \\
\zeta_R (s) &= \sum_{n=1}^{+\infty} \frac{1}{n^s}
\end{align}
The prime means that the term with all $n_i = 0$ should be
excluded from the sum. The reflection formulas for an analytical
continuation of zeta functions exist:
\begin{align}
&\Gamma \Bigl(\frac{s}{2}\Bigr) \pi^{-s/2} \zeta_R (s) = \Gamma
\Bigl(\frac{1-s}{2}\Bigr) \pi^{(s-1)/2} \zeta_R (1-s)
\label{refl1}
\\ &\Gamma \Bigl(\frac{s}{2}\Bigr) \pi^{-s/2} Z_3 (a,b,c;s) =
(abc)^{-1} \Gamma \Bigl(\frac{3-s}{2}\Bigr) \pi^{(s-3)/2} Z_3
\Bigl(\frac{1}{a},\frac{1}{b},\frac{1}{c}; 3-s \Bigr)
\label{refl2}
\end{align}
 By use of reflection
formulas  (\ref{refl1}), (\ref{refl2}) one gets:
\begin{align}
Z_3 \Bigl(\frac{1}{a}, \frac{1}{b}, \frac{1}{c} ; -1 \Bigr) &= -
\frac{abc}{2\pi^3} Z_3 \Bigl(a,b,c ; 4 \Bigr) \\
\zeta_R (-1) &= -\frac{1}{12} .
\end{align}
The renormalized Casimir energy can therefore be written as:
\begin{equation}
E_{cavity} (a,b,c) = - \frac{abc}{16\pi^2} Z_3(a,b,c; 4) +
\frac{\pi}{48} \Bigl(\frac{1}{a} + \frac{1}{b} + \frac{1}{c} \Bigr)
. \label{f24}
\end{equation}
One can check that the formulas (\ref{f24}) and (\ref{box})
coincide identically and yield the Casimir energy for a perfectly
conducting cavity.

\subsection{Geometric interpretation}
In this section we suggest a geometric interpretation of the main
formulas in terms of geometric optics. This interpretation clarifies
the physical meaning of the results (\ref{wave}) , (\ref{box})
obtained, which is always important for further generalizations in
more complicated cases.

Several geometric approaches - a semiclassical method
\cite{semicl}, a worldline approach \cite{Gies} and a method of
geometric optics \cite{Jaffe} have been introduced recently for
the evaluation of the Casimir energies.  Our formulas
(\ref{wave}), (\ref{box}) yield a simple geometric interpretation
for the Casimir energy of the rectangular cavities in terms of
geometric optics.

Optical contributions to the Green's function of the scalar field
with Dirichlet and Neumann boundary conditions have the form:
\begin{eqnarray}
G_{optical}^{D} ({\bf x},{\bf x^{\prime}},\omega_i ) &=&
\frac{1}{4\pi} \sum_n (-1)^n \sqrt{\Delta_n({\bf x},{\bf
x^{\prime}})} \exp (i \omega_i l_n ({\bf x},{\bf x^{\prime}}) ) \\
G_{optical}^{N} ({\bf x},{\bf x^{\prime}},\omega_i ) &=&
\frac{1}{4\pi} \sum_n \sqrt{\Delta_n({\bf x},{\bf x^{\prime}})} \exp
(i \omega_i l_n ({\bf x},{\bf x^{\prime}}) ) .
\end{eqnarray}
Here $l_n({\bf x},{\bf x^{\prime}})$ is the length of the optical
path that starts from ${\bf x}$ and arrives at ${\bf x^{\prime}}$
after n reflections from the boundary. $\Delta_n({\bf x},{\bf
x^{\prime}})$ is the enlargement factor of classical ray optics. For
planar boundaries it is given by $\Delta_n({\bf x},{\bf x^{\prime}})
= 1/l_n^2$. From (\ref{plates}) it follows that for two parallel
plates the Casimir energy of the electromagnetic field can be
expanded as:
\begin{equation}
\begin{split}
E&= \frac{S}{2\pi^2} \int_{0}^{+\infty} d k \,\, k^2 \,\, \ln
\Bigl(1 - \exp(- 2 k a) \Bigr) = \\
&= - 2aS \int_{0}^{+\infty} dk \frac{4\pi k^2}{(2\pi)^3}
\sum_{n=1}^{+\infty} \frac{\exp(-2ank)}{2an} = - \sum_{\omega_i}
\sum_{n=1}^{+\infty}\frac{\exp(-2an\omega_{i})}{2an}. \label{opt1}
\end{split}
\end{equation}
Here $\sum_{\omega_i}$ is a sum over all photon states (with
frequencies $\omega_{i} = \sqrt{k_x^2+k_y^2+k_z^2}$) in an infinite
space. The righthandsight of (\ref{opt1}) can be written in terms of
optical Green's functions:
\begin{equation}
\sum_{\omega_i} \sum_{n=1}^{+\infty} \frac{\exp(-2an\omega_i)}{2an}
= 2\pi \sum_{\omega_i} (G_{optical}^{D} ({\bf x},{\bf x}, a, i
\omega_i ) + G_{optical}^{N} ({\bf x},{\bf x}, a, i \omega_i ))
\label{f20}
\end{equation}
Note that terms with odd reflections from Dirichlet and Neumann
Green's functions cancel each other due to the factor $(-1)^n$
present in optical Dirichlet Green's function. This is why only
periodic paths with even number of reflections from the boundary
$l_{2n}=2an$ enter into the expression for the Casimir energy.

Now consider the formula for the cavity (\ref{box}) (to obtain the
Casimir energy of a waveguide (\ref{wave}) in terms of optical
Green's functions the arguments are the same, just start from two
parallel plates). Imagine that there is a waveguide with side
lengths $b$ and $c$. In order to obtain the rectangular cavity we
have to insert two perfectly conducting plates with side lengths $b$
and $c$ (and a distance $a$ apart) inside the waveguide. The
eigenfrequencies that existed in a waveguide were equal to
$\omega_{wave} = \sqrt{(\pi n_2/b)^2+(\pi n_3/c)^2 + p^2}$. Only the
photons with frequencies $\omega_{wave}$ existed in a waveguide, and
these photons start interacting with the plates inserted inside a
waveguide. The optical contribution to the Casimir energy arising
from the interaction of these $\omega_{wave}$ photons with inserted
plates is equal to
\begin{multline}
\begin{split}
&- 2\pi \sum_{\omega_{wave}} (G_{optical}^{D} ({\bf x},{\bf x}, a, i
\omega_{wave} ) + G_{optical}^{N} ({\bf x},{\bf
x}, a, i \omega_{wave} )) = \frac{\pi}{48 a} +\\
&+ \frac{1}{4} \sum_{n_2, n_3 = -\infty}^{+\infty}
\int_{-\infty}^{+\infty} \frac{dp}{2\pi} \ln \biggl(1 - \exp
\Bigl[-2 a \sqrt{\Bigl(\frac{\pi n_2}{b}\Bigr)^2 + \Bigl(\frac{\pi
n_3}{c}\Bigr)^2 + p^2} \Bigr]\biggr)  \label{Eattr}
\end{split}
\end{multline}
(this equality is obtained in analogy to (\ref{opt1}) and
(\ref{f20})). After comparison with (\ref{box}) it is
straightforward to rewrite the Casimir energy of the rectangular
cavity in terms of optical contributions :
\begin{multline}
 E_{cavity}(a,b,c) =  \\
 \begin{split}
 &= \frac{\pi}{48b} + \frac{\pi}{48c} - 2\pi \sum_{\omega_i}
 (G_{optical}^{D} ({\bf x},{\bf x}, c, i \omega_i ) + G_{optical}^{N}
({\bf x},{\bf x}, c, i \omega_i )) - \\
&\;\;\;\;\; - 2\pi \sum_{\omega_{plates}} (G_{optical}^{D} ({\bf
x},{\bf x}, b, i \omega_{plates} ) + G_{optical}^{N} ({\bf x},{\bf
x}, b, i \omega_{plates} )) - \\
&\;\;\;\;\; - 2\pi \sum_{\omega_{wave}} (G_{optical}^{D} ({\bf
x},{\bf x}, a, i \omega_{wave} ) + G_{optical}^{N} ({\bf x},{\bf x},
a, i \omega_{wave} )) . \end{split}   \label{main}
\end{multline}
Here we sum over all eigenfrequencies of the electromagnetic field
in $3$ cases: when there is an infinite space ($\omega_i =
\sqrt{k_x^2+k_y^2+k_z^2}$), two parallel plates ($\omega_{plates} =
\sqrt{(\pi n_2/b)^2 + k_x^2 + k_z^2}$) and an infinite waveguide \\
($\omega_{wave}= \sqrt{(\pi n_2/b)^2 + (\pi n_3/c)^2 + (k_x)^2 } $).

The first two terms in (\ref{main}) may have the following geometric
interpretation: from (\ref{f21}) it follows that
\begin{eqnarray}
\frac{\pi}{48 b} &=&  \pi \sum_{\omega_{mw}} (G_{optical}^{D} ({\bf
x},{\bf x}, b, i \omega_{mw} ) +
G_{optical}^{N} ({\bf x},{\bf x}, b, i \omega_{mw} ))  \\
\frac{\pi}{48 c} &=&  \pi \sum_{\omega_{mw}} (G_{optical}^{D} ({\bf
x},{\bf x}, c, i \omega_{mw} ) + G_{optical}^{N} ({\bf x},{\bf x},
c, i \omega_{mw} )) \; ,
\end{eqnarray}
where $\omega_{mw}$ is an eigenfrequency of a main wave in a
waveguide. So it is possible to express Casimir energies of
perfectly conducting rectangular cavities in terms of optical
Green's functions only.

It is interesting that the Casimir energy of a perfectly conducting
cavity can be written in terms of eigenfrequencies of the
electromagnetic field in a free space, between two perfectly
conducting plates and inside a perfectly conducting waveguide.

\subsection{The experiment}
For the experimental check of the Casimir energy for the
rectangular cavity one should measure the force somehow. We think
about the following possibility: one should insert two parallel
perfectly conducting plates inside an infinite perfectly
conducting waveguide and measure the force acting on one of the
plates as it is being moved through the waveguide.  The distance
between the inserted plates is $a$.

To calculate the force on each plate the following gedanken
experiment is useful. Imagine that $4$ parallel plates are
inserted inside an infinite waveguide and then $2$ exterior plates
are moved to spatial infinities. This situation is exactly
equivalent to $3$ perfectly conducting cavities touching each
other. From the energy of this system one has to subtract the
Casimir energy of an infinite waveguide, only then do we obtain
the energy of interaction between the interior parallel plates,
the one that can be measured in the proposed experiment (the
subtraction of the term (\ref{EV2}) is just the same subtraction
for two parallel plates). Doing so we obtain the attractive force
on each interior plate inside the waveguide:
\begin{equation}
F_{attr} (a,b,c) = - \frac{\partial E_{attr}(a,b,c)}{\partial a},
\label{fattr}
\end{equation}
where
\begin{multline}
\begin{split}
&E_{attr} (a,b,c) =  \frac{\pi}{48 a} +
\\ &+ \frac{1}{4} \sum_{n_2, n_3 = -\infty}^{+\infty}
\int_{-\infty}^{+\infty} \frac{dp}{2\pi} \ln \biggl(1 - \exp
\Bigl[-2 a \sqrt{\Bigl(\frac{\pi n_2}{b}\Bigr)^2 + \Bigl(\frac{\pi
n_3}{c}\Bigr)^2 + p^2} \Bigr]\biggr)  \label{attr}
\end{split}
\end{multline}
coincides with (\ref{Eattr}). We note that our formula
(\ref{fattr}) for the special case $b=c$ coincides with the
formula (6) in reference \cite{Jaffe2}, there it was obtained
using a different method and presented in a different mathematical
form.

To obtain the energy of interaction between the opposite sides of a
{\it single} cavity one should subtract from the expression
(\ref{box}) the Casimir energy of the same box without these two
sides, i.e. one has to subtract from (\ref{box}) the expression for
the Casimir energy of a waveguide of a finite length. To our
knowledge the expression for the Casimir energy of a finite length
waveguide is not known up to now.

It was often argued that the constant repulsive force (for a fixed
cross section) derived from (\ref{box}) can be measured in
experiment. However, without the subtraction just mentioned it is
not possible to measure the forces in any realistic experiment,
this is why it is not possible to use the expression (\ref{box})
directly to calculate the force in the experiments.  However, it
can be used to derive a measurable in experiments expression for
the force between the parallel plates inserted inside an infinite
waveguide of the same cross section as the plates.

 Using the same technique as before it is possible to
 generalize our formulas (\ref{fattr}), (\ref{attr})
 for the case of an  infinite waveguide with an arbitrary
 cross section. The force between the  two plates inside
 this waveguide can be immediately written :
\begin{align}
F (a) &= - \frac{\partial E_{arb}(a)}{\partial a}, \\
E_{arb} (a) &= \sum_{\omega_{wave}} \frac{1}{2} \ln (1-\exp(-2 a
\, \omega_{wave})), \label{r7}
\end{align}
the sum here is over all  TE and TM eigenfrequencies
$\omega_{wave}$ for the waveguide with an arbitrary cross section
and an infinite length. Thus it can be said that {\it the exchange
of photons with the eigenfrequencies of a waveguide between the
inserted plates always yields the attractive force between the
plates}.

To get the free energy $F_{arb} (a,\beta)$ for bosons
 at nonzero temperatures $\beta=1/T$ one has to make the
substitutions (see Sec. $2.3$, the formula (\ref{m2})):
\begin{align}
 p &\to p_m = \frac{2\pi m}{\beta} , \\
 \int_{-\infty}^{+\infty}  \frac{dp}{2\pi} &\to \frac{1}{\beta}
 \sum_{m=-\infty}^{+\infty}.
\end{align}
Thus the free energy describing the interaction of the two
parallel perfectly conducting plates inside an infinite perfectly
conducting waveguide of an arbitrary cross section has the form:
\begin{align}
F_{arb} (a,\beta) = &\frac{1}{\beta} \sum_{\lambda_{k D}}
\sum_{m=-\infty}^{+\infty} \: \frac{1}{2} \ln \Bigl(1-\exp
(-2a\sqrt{\lambda_{k D}^2 + p_m^2} ) \Bigr) + \nonumber \\ + &
\frac{1}{\beta} \sum_{\lambda_{i Neum}} \sum_{m=-\infty}^{+\infty}
\: \frac{1}{2} \ln \Bigl(1-\exp (-2a\sqrt{\lambda_{i Neum}^2 +
p_m^2} ) \Bigr) , \label{r20}
\end{align}
where $\lambda^2_{k D}$ and $\lambda^2_{i Neum}$ are eigenvalues
of the two-dimensional Dirichlet and Neumann problems (a boundary
here coincides with the boundary of each plate inside the
waveguide):
\begin{align}
& \Delta^{(2)} f_k (x,y) = - \lambda_{k D}^2 f_k (x,y) \\
&  f_k (x,y) |_{\partial M} = 0 ,
\end{align}
\begin{align}
& \Delta^{(2)} g_i (x,y) = - \lambda_{i Neum}^2 g_i (x,y) \\
& \frac{\partial g_i (x,y)}{\partial n} \Bigl|_{\partial M} = 0 .
\end{align}

The attractive force between the plates inside an infinite
waveguide of the same cross section at nonzero temperatures is
given by:
\begin{align}
F (a, \beta) &= -\frac{\partial F_{arb} (a, \beta) }{\partial a} =
\nonumber
 \\ & - \frac{1}{\beta} \sum_{\omega_{TD}}
\frac{\omega_{TD}}{\exp(2a\omega_{TD}) - 1} - \frac{1}{\beta}
\sum_{\omega_{TN}} \frac{\omega_{TN}}{\exp(2a\omega_{TN}) - 1}.
\end{align}
Here $\omega_{TD} = \sqrt{p_m^2 + \lambda^2_{k D}} $ and
$\omega_{TN} = \sqrt{p_m^2 + \lambda^2_{i Neum}} $.

The proof of these results will be presented elsewhere.

%%%%%%%%%%%%%%%%%%%%%%%%%%%%%%%%%%%%%%%%%%%%%%%%
%% BACKMATTER
%%%%%%%%%%%%%%%%%%%%%%%%%%%%%%%%%%%%%%%%%%%%%%%%

\section*{Acknowledgements}
%\begin{theacknowledgments}

  V.M. thanks D.V.Vassilevich, Yu.V.Novozhilov and V.Yu.Novozhilov
  for suggestions during the preparation of the paper.
  V.M. thanks R.L.Jaffe and M.P.Hertzberg
  for correspondence and discussions. This work has been
  supported in part by a grant RNP $2.1.1.1112$.

%\end{theacknowledgments}

%% If the bibliography is produced without BibTeX comment out the

\bibliographystyle{aipprocl} % if natbib is missing


%  \typeout{******************************************}
%  \typeout{** Please run "bibtex \jobname" to optain}
%  \typeout{** the bibliography and then re-run LaTeX}
%  \typeout{** twice to fix the references!}
%  \typeout{******************************************}

\end{document}